\documentclass[11pt,a4paper]{article}

\usepackage[T1]{fontenc}
\usepackage[utf8]{inputenc}
\usepackage{lmodern}

\usepackage[english]{babel}
\usepackage{csquotes}

\usepackage[a4paper,margin=2.5cm]{geometry}

\usepackage{amsmath}
\usepackage{amssymb}
\usepackage{amsfonts}
\usepackage{bm}
\usepackage{mathtools}

\usepackage{graphicx}
\usepackage{float}
\usepackage{subcaption}
\usepackage{booktabs}

\usepackage{xcolor}
\usepackage[
  colorlinks=true,
  linkcolor=black,
  citecolor=blue,
  urlcolor=blue,
  pdfborder={0 0 0}
]{hyperref}

\usepackage[nameinlink,capitalise]{cleveref}

\let\cref\Cref

\crefname{figure}{Figure}{Figures}
\crefname{table}{Table}{Tables}
\crefname{section}{Section}{Sections}
\crefname{equation}{Equation}{Equations}
\crefname{appendix}{Appendix}{Appendices}

\usepackage[
  backend=bibtex,
  style=numeric-comp,
  sorting=none,
  maxbibnames=99,
  giveninits=true,
  doi=true,
  isbn=false,
  url=true
]{biblatex}

\DeclareFieldFormat[patent]{date}{Priority date\addcolon\space#1}

\renewbibmacro*{patent:date}{%
  \printdate
}

\usepackage{array}
\usepackage{multirow}
\usepackage{enumitem}
\setlist[itemize]{label=--}
\usepackage{algorithm}
\usepackage{algorithmic}

\usepackage{listings}

\title{Gabor Holography Reinvented}

\author{
  Jesper Glückstad\\[0.5em]
  \small SDU Centre for Photonics Engineering\\
  \small University of Southern Denmark \\
  \small Campusvej 55, 5230 Odense M \\
  \small Denmark
}

\date{\today}

\begin{document}

\maketitle

\begin{abstract}
\noindent
This paper presents a “reinvention” of Gabor Holography that does not suffer optically from the inherent twin-image problem originating back to Gabor's original Nobel Prize awarded invention. 
In-line or on-axis holography was ironically abandoned by its inventor Dennis Gabor himself and was effectively completely \enquote{re-placed} by so-called off-axis holography at the time when Gabor received the Nobel Prize in Physics in 1971. 
However, Gabor Holography is today the method of choice in modern digital holography due to its inherent on-axis, common-path robustness, lower requirements to resolution of the image sensor (or recording material), shorter exposure time, relaxed mechanical stability and temporal coherence requirements. 
However, it still inherently suffers from the aforementioned twin-image problem and, hence, one will find an abundance of papers trying to overcome this challenge by iterative phase retrieval or machine learning based approaches. Gabor Holography Reinvented overcomes this long-lasting twin-image problem for the first time by optical means \cite{gluckstad_2026_ds_patent}.
\end{abstract}

\section{Introduction}
\label{sec:introduction}

Holography has found uses in numerous areas of academia and industry \cite{gluckstad_new_2023,gluckstad_gabor-type_2024,gabor_new_1948,madsen_-axis_2023,pacheco_adaptive_2022,ren_automatic_2019,brault_automatic_2022,oconnor_deep_2020,descoteaux_efficient_2017,huang_holographic_2021,denis_inline_2009,guo_lensfree_2022,ma_quantitative_2021,castaneda_video-rate_2021,curtis_dynamic_2002,madsen_holotile_2022,gluckstad_holotile_2024-1,madsen_axial_2025,madsen_holotile_2025, alvarez-castano_holographic_2025,madsen_lensless_2025,madsen_holovam_2026,banas_holo-gpc_2017,banas_light_2017,gluckstad_holographic_2022,gluckstad_holographic_2023}, including microscopy, light-shaping, particle trapping and manipulation, cryptography, quantum, etc.
Interestingly, while holography’s foundational principles were laid by Nobel laurate, Dennis Gabor \cite{gabor_nobel_1971}, he faced the persistent challenge of the so-called twin-image problem. 
At the time, there was no analytical solution to quantify or eliminate this twin-image issue, leading Gabor to effectively abandon his on-axis holography research after exploring various optical setups. 
This decision stands in contrast to the subsequent proliferation and significance of holography in diverse fields. 
The ability to regain both amplitude and phase information of an optical wavefront after its recording by an intensity-sensitive detector is a key factor of its popularity \cite{gluckstad_new_2023,gluckstad_gabor-type_2024,madsen_-axis_2023}. 
The holographic equation, which reveals this property of holographic image capture, can be written as the intensity of the superposition of two incident wavefronts -- the object $O$ and reference $R$. 
The interference pattern recorded by the image sensor can thus be described by the intensity distribution $I(x,y)$ \cite{goodman_introduction_2017}:

\begin{equation}
	I(x, y)=(O+R)(O+R)^*=O O^*+R R^*+O R^*+O^* R
\end{equation}
where $O$ is the complex amplitude distribution of the object light, i.e., the scattered light from the object generated when the object is illuminated by the light, and $R$ is the complex amplitude distribution of the reference light, i.e., the unobstructed light from the light beam. 
Spatial variables $x$ and $y$ denote the $x$- and $y$-coordinates on the image sensor and therefore correspond to the discretized pixel coordinates and $^*$ denotes the complex conjugate. The $x$- and $y$-coordinates of $O$ and $R$ are omitted for convenience.

The four terms comprising the holographic equation are then:
\begin{itemize}
	\item $OO^*$; the scattered object wave interfering with itself, thus denoted here as “self-interference” term,
	\item $RR^*$; the directly transmitted and un-scattered reference wave corresponding to the background, and
	\item $OR^*$ and $O^*R$; real and virtual images, respectively, of the original object wavefront, scaled by the reference wave, containing both amplitude and phase information of the object.
\end{itemize}

Typically, due to the complex conjugation, the term $O^*R$ is often denoted as the so-called twin-image \cite{elena_stoykova_twin-image_2014} of the object. 
Now, to extract the amplitude and phase of the original object wavefront, the term $OR^*$ -- i.e., the real image, must be isolated to the greatest extent.

In off-axis holography, in which the illuminating light is split into an object beam and an angled reference beam, thus introducing a spatial carrier wave, the four terms of the holographic equation above can \enquote{simply} be spatially separated and filtered-out in Fourier space \cite{leith_reconstructed_1962,leith_wavefront_1963,leith_wavefront_1964,pi_review_cgh_2022}. 
However, when discussing on-axis holography, the object illuminating light source serves as both the object and reference wave. It is conceptually divided into the two waves, but originates from the same light beam direction.
Thus, both wavefronts propagate axially in line, not allowing for spatial separation of the terms. 
It should be noted that on-axis holography can also be embodied by using an independent reference wave -- again taken from the same light source due to holographic coherence requirements -- co-propagating in line with the object wave. 
This is particularly relevant when the object is semi-opaque, opaque or holographically illuminated in a reflection geometry.

The reconstruction of the object beam in on-axis holography is typically undertaken with iterative numerical methods, usually from derivations of the well-known Gerchberg-Saxton algorithm \cite{gerchberg_practical_1972,madsen_comparison_2022,pang_speckle-reduced_2019,liu_symmetrical_2006,wu_adaptive_2021,momey_fienups_2019}. 
While good results can be achieved, these iterative phase retrieval algorithms are inherently non-convex, not ensuring convergence to the global minimum. 
In addition, the iterative nature limits the reconstruction speed for most practical applications.

In the last decade, machine learning algorithms using e.g., deep neural networks are also tackling the same problem of wavefront retrieval and is showing promising results \cite{huang_holographic_2021,madsen_-axis_2023,sinha_lensless_2017,ren_end--end_2019,rivenson_phase_2018,ju_learning-based_2022}. 
While inference can be performed at high reconstruction speed, the training of the deep neural networks requires both long training time and many thousand training samples.

On 16 Aug. 2024, I filed a +50 pages patent application \cite{gluckstad_2026_ds_patent} on a method that fully overcomes -- optically -- this long-lasting challenge of removing the disturbing twin-image in Gabor Holography. 
In-line or on-axis holography was ironically abandoned by its inventor Dennis Gabor himself and was effectively \enquote{re-placed} by off-axis holography when Gabor received the Nobel Prize in Physics in 1971.
However, Gabor Holography is today the method of choice in modern digital holography due to its inherent on-axis, common-path robustness, lower requirements to resolution of the image sensor (or recording material), shorter exposure time, relaxed mechanical stability and temporal coherence requirements. 
Gabor Holography Reinvented overcomes this long-lasting twin-image problem for the first time by optical means.

\section{Challenge}
In general, we have that the acquired intensity, $I$, of the holographic recording is given by:
\begin{equation}
	I=1+o \otimes h+(o \otimes h)^*+|o \otimes h|^2
\end{equation}

\begin{figure}[h!]
	\centering
	\includegraphics[width=\textwidth]{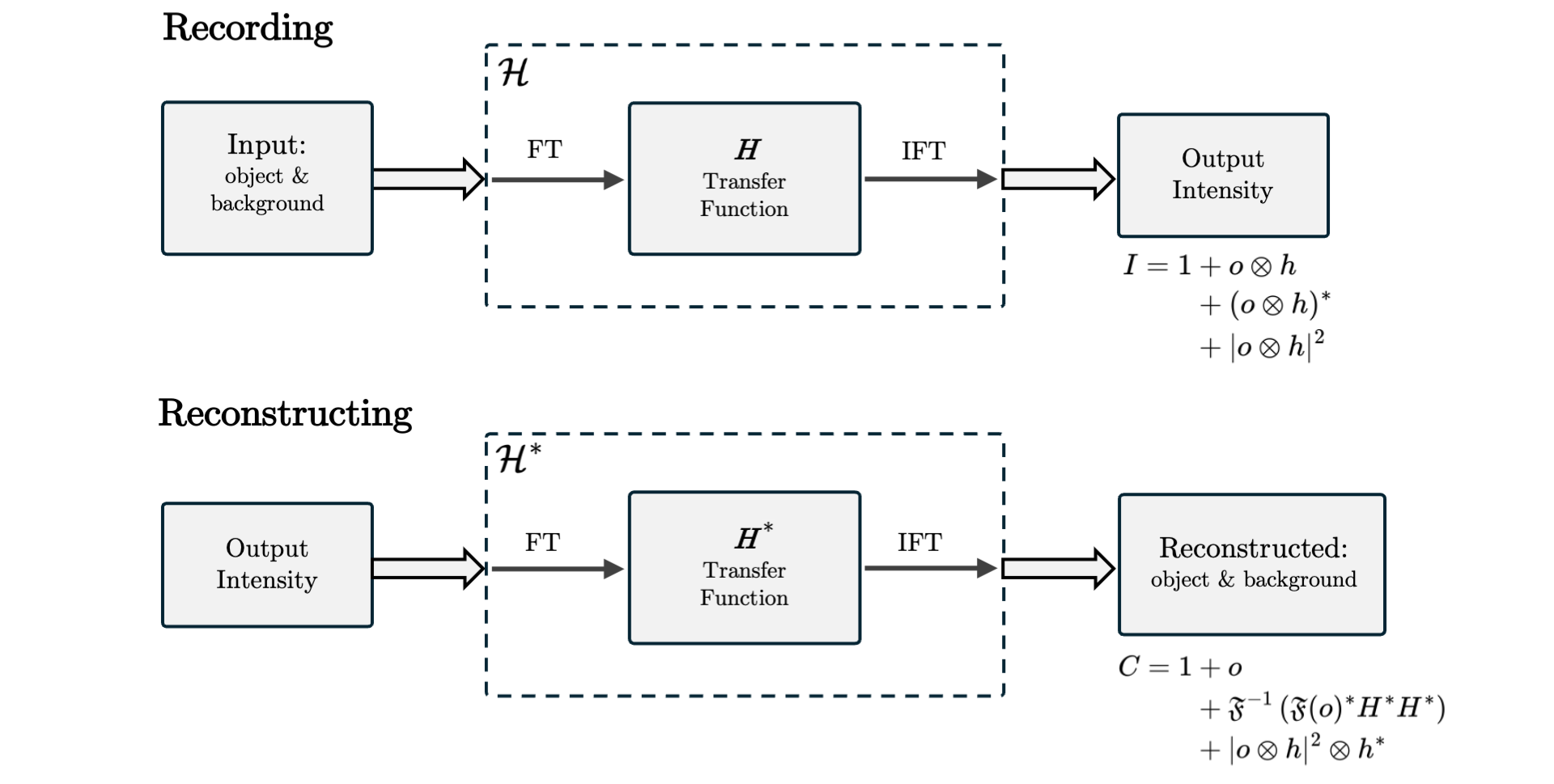}
	\caption{Generalized recording and reconstruction process of holography. $\mathcal{H}$ and its conjugate $\mathcal{H}^*$ describe the optical system.}
	\label{fig:process}
\end{figure}

Referring to \cref{fig:process}, the reconstruction, $C$, is given by:
\begin{equation}
	C=1+o+\mathfrak{F}^{-1}\left(\mathfrak{F}(o)^* H^* H^*\right)+|o \otimes h|^2 \otimes h^*
\end{equation}

The aim is now to identify a Transfer Function, $H$, such that we minimize the influence of this term:
\begin{equation}
\label{eq:min}
	\operatorname{\mathbf{Min}}\left[\mathfrak{F}^{-1}\left(\mathfrak{F}(o)^* H^* H^*\right)+|o \otimes h|^2 \otimes h^*\right]
\end{equation}
\textit{while} we seek to maximize the influence of the reconstructed object on its background:
\begin{equation}
\label{eq:max}
	\operatorname{\mathbf{Max}}\left[ 1+o \right]
\end{equation}

\begin{figure}[t]
	\centering
	\includegraphics[width=\textwidth]{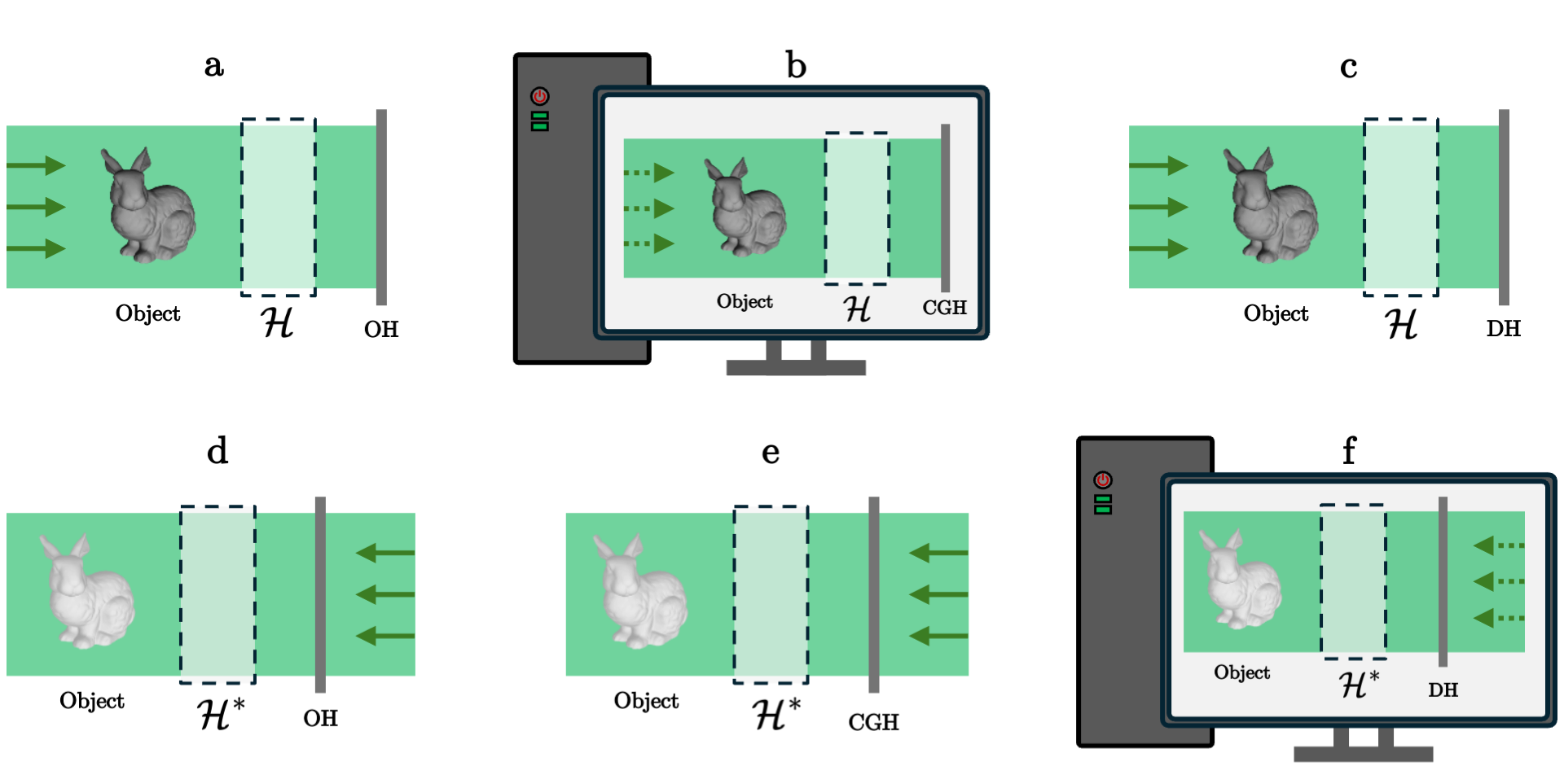}
	\caption{Schematic of hologram recording and reconstruction processes. (a, d) Optical holography (OH), (b, e) computer-generated holography (CGH), (c, f) digital holography (DH). $\mathcal{H}$ and its conjugate $\mathcal{H}^*$ describe any operation between the input object and its output.}
	\label{fig:schematic}
\end{figure}

Using the principle of the stationary phase method we can find the local spatial frequency of an optical wavefront modulating element (ideally a lossless \enquote{phase shaper}) as a spatial derivative of its spatial phase in cartesian or polar coordinates:
\begin{align}
	f_x = \frac{1}{2\pi} \frac{\partial \varphi(x, y)}{\partial x} \quad f_y = \frac{1}{2\pi} \frac{\partial \varphi(x, y)}{\partial y} \\
	f_r = \frac{1}{2\pi} \frac{\partial \varphi(r, \theta)}{\partial r} \quad f_\theta = \frac{1}{2\pi} \frac{\partial \varphi(r, \theta)}{\partial \theta}	
\end{align}
Using these expressions makes it possible to design various Transfer Functions, $H$, that use phase-only, lossless modulation to engineer the minimization of \cref{eq:min} while maximizing \cref{eq:max}. The overall system operator describing the transform from input to output field, $\mathcal{H}$, and its constituent transfer function $H$, thus encapsulates all modalities, including, but not limited to, free-space propagation, optical phase shapers, and any combinations of these. In \cref{fig:schematic}, the various instantiations of in-line holography are shown, including all-optical holography, computer-generated holography, and digital holography. In all cases, with knowledge of the Transfer Function and its complex conjugate, $H^*$, the object can be reconstructed.

In the following section, a simple, but not trivial, example of such a Transfer Function, an axicon phase shaper inserted in the spatial frequency domain, is given. Following a derivation of the resulting image sensor read-out, I show how the effect of the axicon phase, as well as knowledge of the conjugate Transfer Function, enables an object reconstruction with far more phase-accuracy than conventional free-space propagation holography.

\section{Axicon-based Phase Shaper}
A diffractive axicon with radial grating period, $d$, and center phase, $\varphi_0$, can be expressed by:
\begin{equation}
	H=\exp \left(i \varphi_0\right) \exp \left(-i 2 \pi \frac{r^\prime}{d}\right)=\exp \left(i \varphi_0\right) \exp \left(-i 2 \pi \frac{r_0 r^\prime}{\lambda f}\right)
\end{equation}
where we have the relation between the annular radius, $r_0$, and the radial grating period, $d$:
\begin{equation}
	r_0 = \frac{\lambda f}{d}
\end{equation}

The conjugated diffractive axicon is given by:
\begin{equation}
	H^*=\exp \left(-i \varphi_0\right) \exp \left(i 2 \pi \frac{r_0 r^\prime}{\lambda f}\right)
\end{equation}

\begin{figure}[tb]
	\centering
	\includegraphics[width=.97\textwidth]{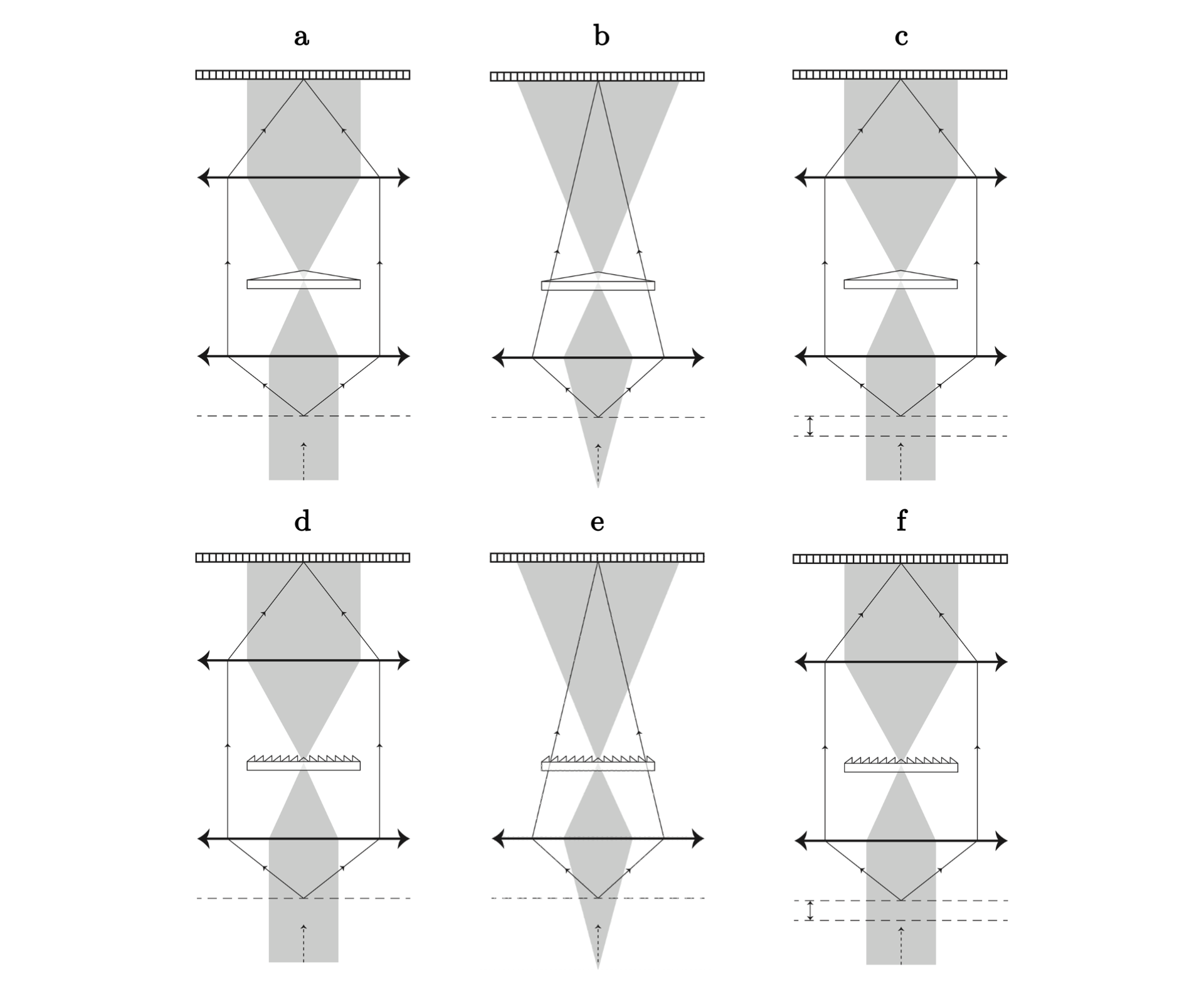}
	\caption{Various axicon modalities. Top row: refractive axicons, bottom row: diffractive axicons. Left column: 4F-based, center column: single imaging lens, right column: combination of free-space propagation and axicon modulation.}
	\label{fig:modalities}
\end{figure}

\begin{figure}[htb]
	\centering
	\includegraphics[width=.9\textwidth]{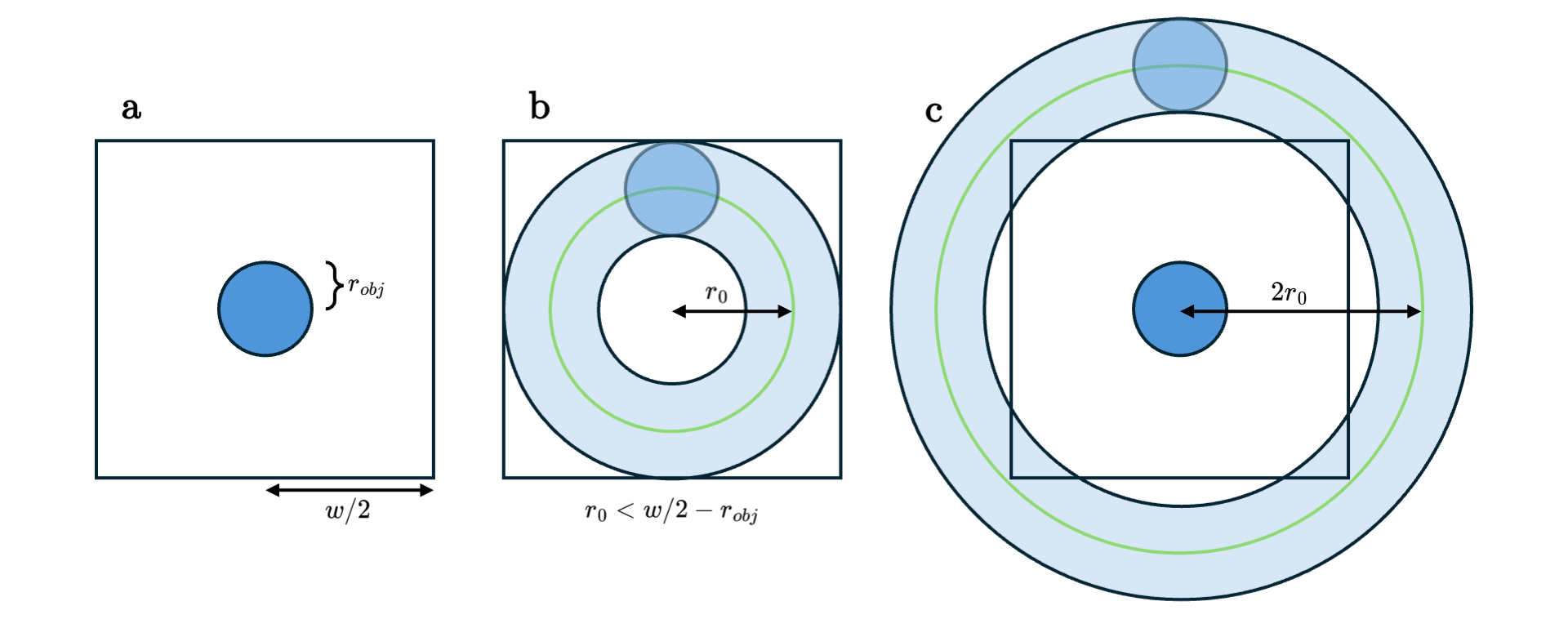}
	\caption{Image sensor read-out and reconstruction. (a) Object imaged without axicon placed in spatial frequency domain, (b) with axicon placed in spatial frequency domain. Holographic terms co-located in bands of characteristic width $r_{obj}$, (c) reconstructed object with twin-image term in band with radius $2r_0$.}
	\label{fig:axicon-ring}
\end{figure}

Giving rise to the corresponding conjugate convolution kernel:
\begin{equation}
	h^* \simeq \exp \left(-i \varphi_0\right) \delta\left(r-r_0\right)
\end{equation}
Image sensor read-out can be derived as:
\begin{equation}
\label{eq:ring-colocation}
\begin{gathered}
I=H(0,0)+\left(o^* \otimes h^*\right)^*+\left(o^* \otimes h^*\right)+\left|o^* \otimes h^*\right|^2 \Rightarrow \\
I=\exp \left(i \varphi_0\right)+\exp \left(i \varphi_0\right)\left(o^* \otimes \delta\left(r-r_0\right)\right)^*+\exp \left(-i \varphi_0\right) o^* \otimes \delta\left(r-r_0\right)+\left|o^* \otimes \delta\left(r-r_0\right)\right|^2
\end{gathered}
\end{equation}
\cref{eq:ring-colocation} describes the image sensor read-out given an axicon placed in the spatial frequency domain, as shown in \cref{fig:modalities}d. As seen in \cref{fig:axicon-ring}b, an object with characteristic radius $r_{obj}$ naturally limits the annular radius, $r_0$, of the axicon. Due to the convolution with the kernel $h^*$, the object information contained in the holographic terms of \cref{eq:ring-colocation} is contained within a set of \enquote{co-located} circular bands of width $r_{obj}$. To maintain all necessary information to recover the original object, the following limit must be enforced:
\begin{equation}
	r_0 < w/2 - r_{obj}
\end{equation}

Re-calling terms from the reconstruction:
\begin{equation}
\begin{aligned}
& C=1+o+(o \otimes h)^* \otimes h^*+|o \otimes h|^2 \otimes h^* \Rightarrow \\
& C=1+o+o^* \otimes\left(h^* \otimes h^*\right)+\left|o^* \otimes h^*\right|^2 \otimes h^*
\end{aligned}
\end{equation}
where inserting $h^*$ gives:
\begin{equation}
\label{eq:reconstruction-total}
	C = 1+o + \exp \left( -2i\varphi_0 \right)o^* \otimes \delta (r-2r_0) + \exp \left( -i\varphi_0 \right) |o^* \otimes \delta(r-r_0)|^2 \otimes \delta(r-r_0)
\end{equation}
which can be simplified to:
\begin{equation}
	C = 1+o + \frac{1}{2} \exp\left( -i\varphi_0 \right) |o|^2 + \left[ \exp\left( -2i\varphi_0 \right) o^* + \frac{1}{2} \exp\left(-i\varphi_0\right)|o|^2 \right]\otimes \delta(r-2r_0)
\end{equation}
Alternatively, by using the original axicon (e.g., in hardware) one can recreate the conjugate object:
\begin{equation}
\label{eq:reconstruction-total-conjugate}
	\tilde{C} = 1 + o^* + \frac{1}{2} \exp \left( i\varphi_0 \right)|o|^2 + \left[ \exp\left( 2i\varphi_0 \right) o + \frac{1}{2} \exp\left( i\varphi_0 \right) |o|^2 \right] \otimes \delta (r-2r_0)
\end{equation}
Importantly, due to the repeated convolution with $h^*$, the twin-image terms in \cref{eq:reconstruction-total,eq:reconstruction-total-conjugate} are now pushed into bands of radius $2r_0$. Hence, the twin image is spatially separated from the object reconstruction through optics alone, as is visualized in \cref{fig:axicon-ring}c, and will not meaningfully interfere with the object reconstruction. Therefore, discarding all terms convolved with the ring-shaped localization function $\delta(r-2r_0)$, we have:
\begin{align}
	\label{eq:Co}
	C_o &= 1 + o + \frac{1}{2} \exp\left(-i\varphi_0 \right)|o|^2 \\
	\label{eq:CoConj}
	\tilde{C}_o &= 1 + o^* + \frac{1}{2} \exp \left( i\varphi_0 \right) |o|^2
\end{align}
Yielding the reconstruction of the object $o$, or its conjugate $o^*$, only affected by a constant offset and the $50\%$ reduced self-interference term. 

The result of \cref{eq:Co,eq:CoConj} is a key insight, in that it is agnostic to the particular phase shaper, as long as the twin-image is pushed outside the signal area. Hence, this result can be generalized by designing a phase shaper that generates the desired parametric kernel, given in polar coordinates with angular coordinate $\psi$:
\begin{equation}
\label{eq:generalized-h}
	h_p^* \simeq \delta(r - r_0(\psi))
\end{equation}
For simplicity in derivation, the phase shaper was chosen as an axicon, resulting in a ring-shaped kernel. However, one can imagine specially designed phase shapers optimized for different properties of the reconstruction, e.g., size of the signal area.

\newpage
\section{Phase-only Object}
With the acquired result of \cref{eq:Co,eq:CoConj}, applicable for all $h_p^*$ designed such that the twin-image is pushed out of the signal area, it is worthwhile to consider the special case of a phase-only object, i.e., an object that is purely phase-modulating. Phase-only objects have e.g., $|o| = 1$ so that we have the substantially simplified scenario:
\begin{align}
	C_{po} &= 1 + \left[o + \frac{1}{2} \exp \left(-i\varphi_0\right)\right]_{obj} = 1 + \left[\exp \left( i \phi_{obj} \right) + \frac{1}{2} \exp \left( -i \varphi_0 \right) \right]_{obj} \\
	\tilde{C}_{po} &= 1 + \left[o^* + \frac{1}{2} \exp \left(i\varphi_0\right)\right]_{obj} = 1 + \left[\exp \left( -i \phi_{obj} \right) + \frac{1}{2} \exp \left( i \varphi_0 \right) \right]_{obj}
\end{align}
which will be even simpler for an applied phase shaper center-phase equal to zero:
\begin{align}
	C^\prime_{po} &=  1 + \left[\exp \left( i \phi_{obj} \right) + \frac{1}{2} \right]_{obj} \\
	\tilde{C}^\prime_{po} &=  1 + \left[\exp \left( -i \phi_{obj} \right) + \frac{1}{2} \right]_{obj} 
\end{align}
Hence, within object boundaries, we can reconstruct a perfectly correct phase object superposed the constant $\frac{3}{2}$.

\section{Free-space Propagated Object}
If the object terms, $o$ or $o^*$, are themselves a result of a quadratic-phase convolved free-space propagation operation, as illustrated in \cref{fig:modalities}c and f, with kernel, $f$, where $o = \hat{o}\otimes f$, the influence of the quadratic self-interference term can be further reduced. 
In that case, we have the back-propagated reconstructions:
\begin{align}
	C_{fo} &= 1 + \hat{o} + \frac{1}{2} \exp \left( -i\varphi_0 \right) |\hat{o}\otimes f|^2 \otimes f^* \\
	\tilde{C}_{fo} &= 1 + \hat{o}^* + \frac{1}{2} \exp \left( i\varphi_0 \right) |\hat{o}\otimes f|^2 \otimes f^*
\end{align}

The self-interference term is often considered negligible in conventional Gabor Holography, as it is much less influential than the twin-image. 
However, now that the twin-image has effectively been removed optically, it remains as the only interfering artifact. 
As in conventional Gabor Holography, the propagation distance directly affects the influence of the self-interference term.
\newpage

\section{Numerical and Experimental Proof-of-Concept}
\begin{figure}[h]
	\centering
	\includegraphics[width=\textwidth]{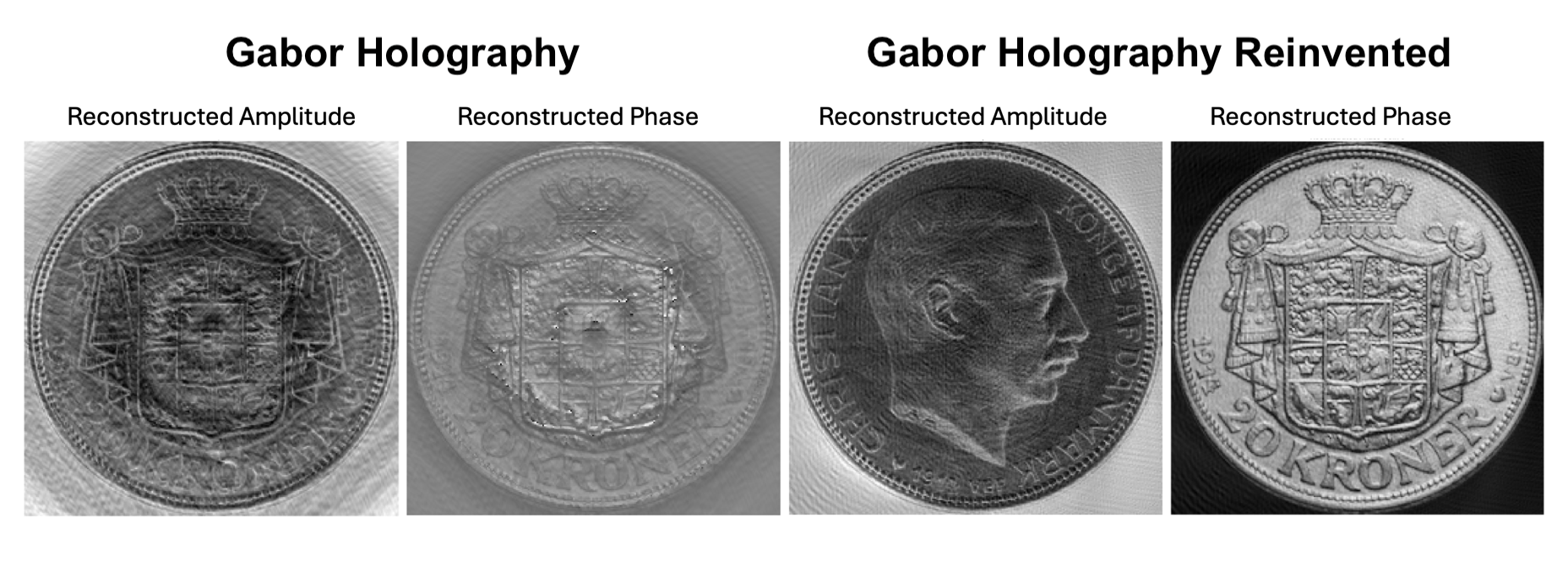}
	\caption{Numerical reconstructions of amplitude and phase of a complex-valued object. The two sides of an old Danish coin are encoded as amplitude and phase, respectively. (Left) reconstruction using Gabor Holography, (right) reconstruction using Gabor Holography Reinvented.}
	\label{fig:comparison}
\end{figure}

In \cref{fig:comparison}, a numerical comparison between Gabor Holography and the Gabor Holography Reinvented described in this paper is shown. 
In this numerical experiment, the two sides of an old Danish coin are encoded as the object amplitude and phase, respectively. In the left case (Gabor Holography), $\mathcal{H}$ consists solely of a free-space propagation, as is common in in-line holography. The reconstructions show the well-known disturbing artefacts of both the twin-image, and the self-interference term. On the right, the reconstruction using the described combined phase shaper and free-space propagation modality is shown. It is immediately apparent that the twin-image no longer interferes significantly with the object term, and that we can achieve far greater reconstruction accuracy.

\begin{figure}
\centering
\begin{subfigure}[b]{\textwidth}
	\centering
  \includegraphics[width=\linewidth]{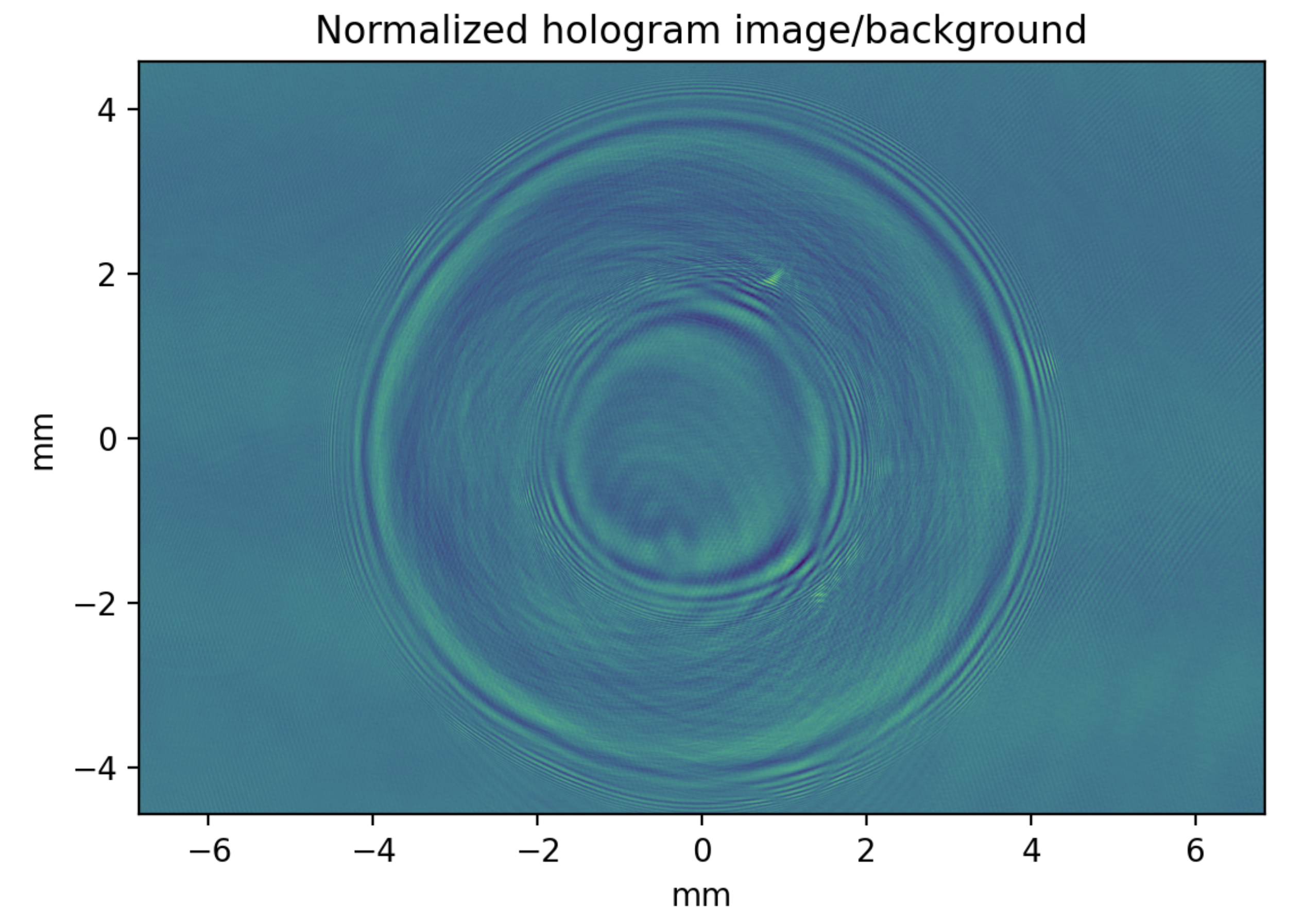}
  \caption{}
  \label{fig:Ng1} 
\end{subfigure}
\begin{subfigure}[b]{\textwidth}
\centering
  \includegraphics[width=1\linewidth]{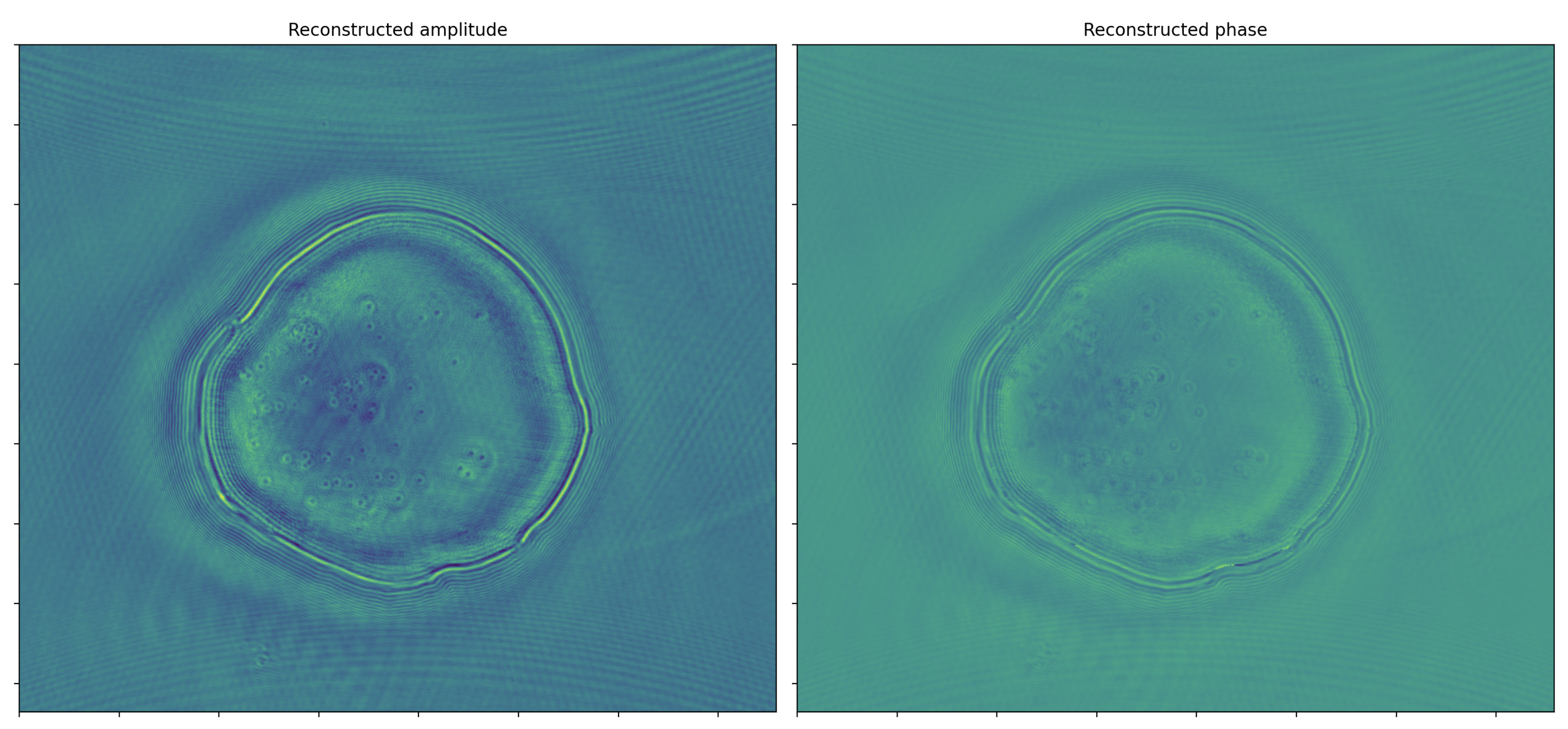}
  \caption{}
  \label{fig:Ng2}
\end{subfigure}
\caption{Experimental verification of Gabor Holography Reinvented. Sensor read-out and reconstruction of water droplet on a microscope slide, using DH as seen in \cref{fig:schematic}c and f. (a) Sensor read-out showing the characteristic band containing the holographic terms. (b) Digitally reconstructed amplitude and phase of the water droplet.}
\label{fig:exp}
\end{figure}

In \cref{fig:exp}, a first proof-of-concept experimental capture and reconstruction for Gabor Holography Reinvented with a refractive axicon phase shaper is presented. The object consists of a water droplet on a microscope slide, observed in a setup conceptually identical to \cref{fig:modalities}a. In \cref{fig:Ng1}, the normalized image sensor read-out is shown. Here, we can clearly see the characteristic band of the holographic terms, corresponding to the convolution with the kernel $h$. \cref{fig:Ng2} shows the reconstructed amplitude and phase of the water droplet. While there is still work needed refining the experimental qualities, as a first proof-of-concept, these results show the effective optical removal of the twin-image term.

\newpage

\section{Reflective 3D Replay with Gabor Holography Reinvented}

\begin{figure}[h]
	\centering
	\includegraphics[width=\textwidth]{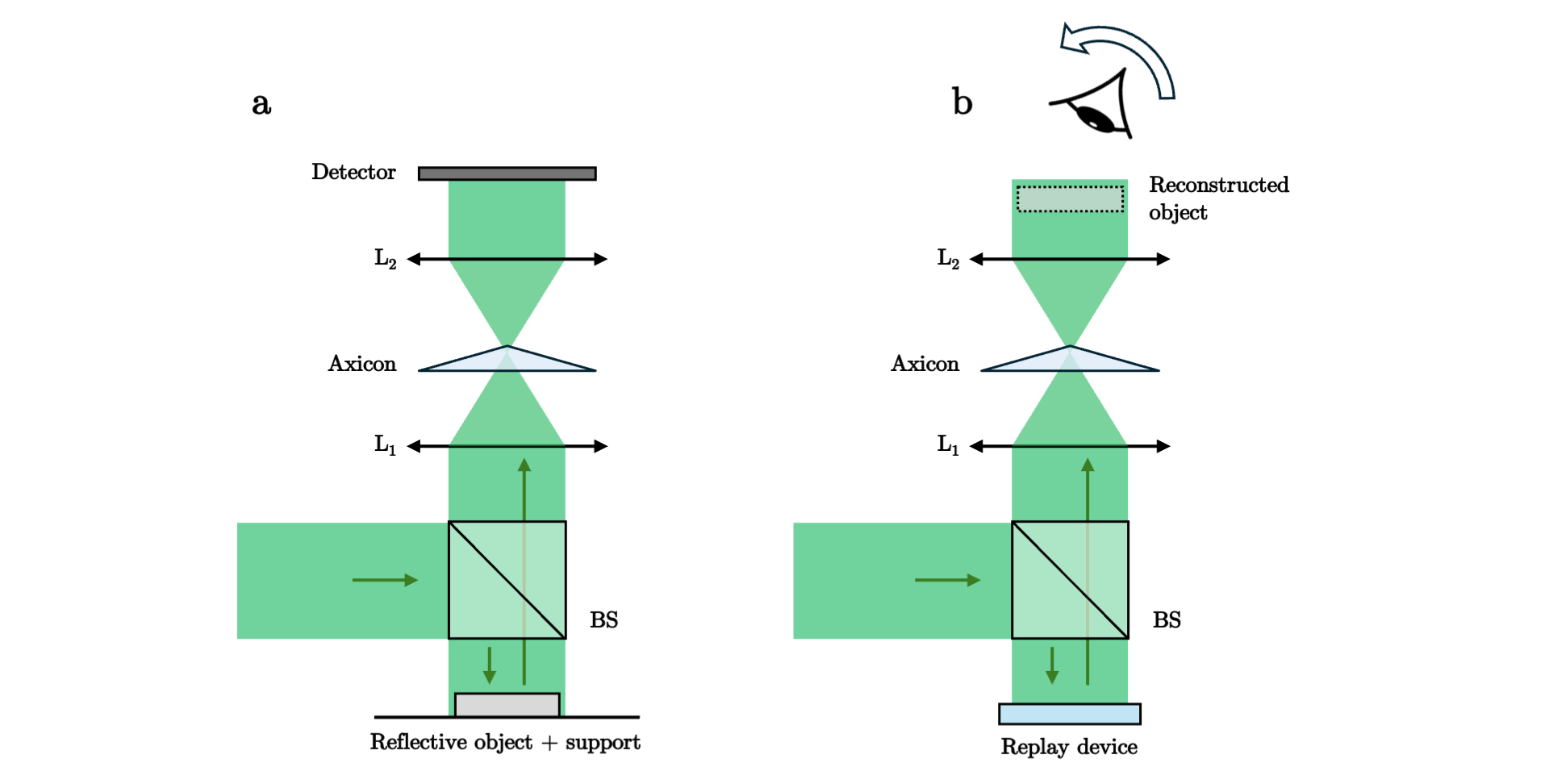}
	\caption{Hardware embodiment example of reflection geometry (left) object + support and (right) analog replay}
	\label{fig:reflective-replay}
\end{figure}

\cref{fig:reflective-replay} shows a hardware embodiment example of Gabor Holography Reinvented in a reflection geometry with analog optical recording (left) and analog optical reconstruction (right). The reflecting object is having a larger reflecting support for generating the in-line reference light. The phase shaper is chosen as standard axicon in both the recording arm and the replay arm of this example setup. The image sensor intensity is sent to a reflecting spatial light modulator for reconstruction. Since a standard axicon is chosen for both recording and reconstruction the setup will replay the complex conjugated object for the observing eye.

\section{Acknowledgement}
I would like to sincerely thank Dr. Andreas Erik Gejl Madsen for his close and constructive collaboration, for carrying out the proof-of-principle experimental work essential to the successful validation of Gabor Holography Reinvented, and for his outstanding support in writing, editing, illustrations, and typesetting the manuscript in LaTeX.

\newpage

\printbibliography

\end{document}